\documentclass[aps,prb,twocolumn,notitlepage,showpacs,superscriptaddress,10pt]{revtex4-2}%
\usepackage{graphicx}
\usepackage{amsmath}
\usepackage{amssymb}
\usepackage{color}
\usepackage{amsfonts}%
\usepackage[caption=false]{subfig}

\usepackage{color}
\usepackage[colorlinks,bookmarks=false,citecolor=darkblue,linkcolor=red,urlcolor=blue]{hyperref}

\definecolor{darkred}{rgb}{0.7,0.0,0.0}

\definecolor{darkblue}{rgb}{0,0.02,0.45}

\definecolor{darkgreen}{rgb}{0.02,0.45,0.0}

\definecolor{violet}{rgb}{0.8,0.2,0.6}

\providecommand{\U}[1]{\protect\rule{.1in}{.1in}}
\newcommand{\ALIO}{Ag$_3$LiIr$_2$O$_6$}
\newcommand{\HLIO}{H$_3$LiIr$_2$O$_6$}

\newcommand{\LIO}{$\alpha$-Li$_2$IrO$_3$}
\newcommand{\rucl}{$\alpha$-RuCl$_3$}
\newcommand{\NIO}{Na$_2$IrO$_3$}

\begin{document}
\title{Role of disorder on the electronic and magnetic properties of {\ALIO}}

\author{Ying Li}\thanks{yingli1227@xjtu.edu.cn}
\affiliation{Department of Applied Physics and MOE Key Laboratory for Nonequilibrium Synthesis and Modulation of Condensed Matter, School of Physics, Xi'an Jiaotong University, Xi'an 710049, China}
\author{Roser Valent{\'\i}}\thanks{valenti@itp.uni-frankfurt.de}
\affiliation{Institut f\"ur Theoretische Physik, Goethe-Universit\"at Frankfurt,
Max-von-Laue-Strasse 1, 60438 Frankfurt am Main, Germany}

\date{\today}

\begin{abstract}
The nature of magnetism in the intercalated honeycomb iridate \ALIO~has been a subject of recent intensive
debate, where the  absence or presence
of antiferromagnetic order has been reported to be related to possible structural disorder effects
and, an enhanced Ir-O hybridization and itinerancy with respect to the parent \LIO~ has been suggested as the origin of distinct 
x-ray spectroscopy features. In the present work we investigate the microscopic nature of the electronic and magnetic properties of \ALIO~  via a combination of density functional theory combined  with  exact  diagonalization of {\it ab initio} derived  models  for  various  experimental  and theoretical  structures. We evaluate
two possible scenarios, the itinerant quasimolecular framework (QMO) on the one hand, and the 
 localized relativistic $j_{\rm eff} = 1/2$ and $j_{\rm eff} = 3/2$ picture on the other hand, and find
  that the second description is still viable for this system. 
 We further calculate resonant inelastic x-ray scattering spectra and show that agreement with experimental observations
 can be obtained if the presence of Ag vacancies leading to changes in Ir filling and structural disorder is assumed.
 Finally, we show that the experimentally observed antiferromagnetic spiral
 magnetic order is reproduced by our {\it ab initio} derived magnetic models.
 
\end{abstract}
\maketitle
\par

\section{Introduction}
%honeycomb systems, other systems
Intensive efforts have been devoted to searching material realizations of the Kitaev spin liquid state in the honeycomb lattice with bond-dependent Ising-like nearest-neighbor interactions~\cite{Kitaev2006,Jackeli2009,Chaloupka2013, Witczak-Krempa2014,Rau16,Schaffer2016,WinterReview,Trebst2017,Cao2018}. 
Promising candidates for the Kitaev spin liquid including the layered honeycomb systems \NIO~\cite{Singh2010, Choi2012, Singh2012}, \LIO~\cite{Singh2012,Gretarsson2013,Freund2016}, and \rucl~\cite{Plumb2014, Kim2015, Johnson2015,Banerjee2016, Banerjee2017, Winter2017, Winter2018} order magnetically either in a zigzag structure (\NIO~and \rucl) or in an incommensurate spiral structure (\LIO~\cite{Williams2016}) due to the presence of further non-Kitaev interactions~\cite{Katukuri2014,Rau2014,Winter2016,Natalia2018,Natalia2018b}.
For the latter, attempts have been made to modulate the magnetic interactions
in terms of intercalated H atoms~\cite{o2012production,Bette2017,Kitagawa2018}. For the resulting \HLIO, magnetic susceptibility, specific heat, and nuclear magnetic resonance (NMR) measurements showed no sign of magnetic order down to 0.05 K~\cite{Kitagawa2018}. 
In fact, theoretical studies~\cite{Li2018,Yadav-Hozoi2018h3liir2o6}  indicated that H positions strongly affect the magnetic interactions, and the resulting magnetic models with bond disorder and vacancies were shown to reproduce the experimentally observed low-energy spectrum in the system ~\cite{Knolle2019, Kao2021}.

%Ag3LiIr2O6 system
Recently, a new member of the intercalated honeycomb iridates family \ALIO~has been synthesized by replacing interlayer Li in \LIO~by Ag atoms~\cite{Bahram2019}. Heat capacity and magnetic susceptibility measurements on those
samples suggested \ALIO~to be closer to the Kitaev limit compared to \LIO. However,  by improving the sample quality controlling that Ag doesn't enter the honeycomb layers, a broad peak in the magnetic susceptibility and heat capacity at $T_N$ = 14 K was observed, 
which, together with a sharper downturn in the magnetic susceptibility at
$T_{N2} = 8$ K and the appearance of spontaneous oscillations in  muon spin relaxation ($\mu$SR) measurements, evidenced the presence of long range incommensurate AFM ordering below T$_{N2}$~\cite{Bahrami2021} of the same type as in \LIO. Further, $\mu$SR measurements and density functional theory calculations reveal a low-temperature ordered state with persistent dynamics down to the lowest temperature below 9 K, and detailed $\mu$SR data is consistent with a coexistence of incommensurate Neel and striped environments~\cite{Chakraborty2021}. The different behavior between the two 
\ALIO~samples was also confirmed by NMR observations~\cite{Wang2020}.
In addition, x-ray absorption and resonant inelastic x-ray scattering (RIXS)
measurements on powder samples in Ref.~\cite{Torre2021} suggested an energy spectrum for \ALIO~ compatible
with the assumption of enhanced Ir-O hybridizations.
An aspect to note is that the sample  characterization of 
Ref. \cite{Torre2021}
doesn't exclude a possible Ir charge disbalance of about 0.2$\%$,
which corresponds to the error bar of the energy dispersive x-ray analysis (EDX). Such a charge disbalance 
could be caused by, e.g., the presence of interlayer Ag vacancies. 

\begin{figure}[t]
\includegraphics[angle=0,width=\linewidth]{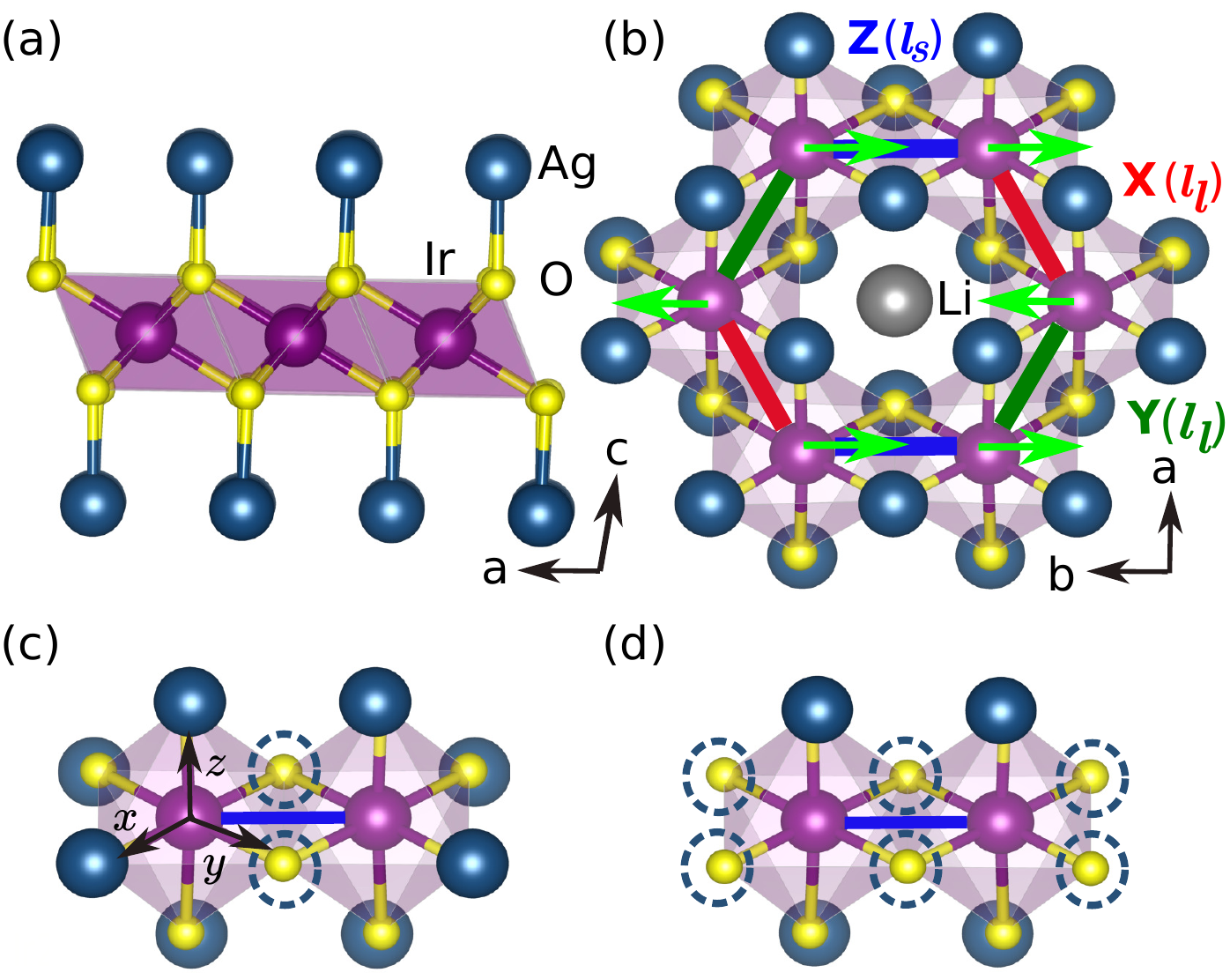}
\caption{Crystal structure of \ALIO in the (a) $ac$ plane and (b) $ab$ plane. Ir, Ag and O are
displayed as magenta, dark blue and yellow balls.  
Red, green, and blue bonds show the three different types of bonds, $X$, $Y$, and $Z$, respectively.
The green arrows in (b) indicate the stripy magnetic configuration used in the calculations. $l_l$ denotes the length of the long bond ($X$ and $Y$) and $l_s$ the length of the short bond ($Z$) for the structure $S_3$ (see the main text).
(c) and (d) show two types of Ag vacancies considered (marked with circles) on
$Z$-bond Ir-Ir clusters corresponding to Ir$^{4.5+}$ and  Ir$^{5+}$, respectively. 
  $x, y, z$ are the Cartesian coordinates for $d$ orbitals. }
\label{fig:structure}
\end{figure}

In view of the above observations, we investigate here the
microscopic origin of the electronic and magnetic properties in \ALIO~in comparison to its parent compound {\LIO} and  analyze the role of disorder effects related to
the presence of Ag vacancies. For that, we perform density functional theory (DFT) calculations combined
with exact diagonalization (ED) of {\it ab initio} based models for various experimental and
theoretically-derived structures.

Such spin models are obtained under the assumption of large spin-orbit coupling and Coulomb repulsion leading to a localized basis of
 $j_{\rm eff}$ = 1/2 and lower lying $j_{\rm eff}$ = 3/2 relativistic orbitals. Alternatively, due to the honeycomb backbone nature of these systems with dominant oxygen assisted $d$-$d$ hybridizations, one can consider a description of the electronic structure in terms of quasimolecular orbitals (QMOs)~\cite{Mazin2012, Foyevtsova2013}.  In the nonrelativistic case, the electronic structure of {\NIO}, for instance, is well described by the QMO basis. The inclusion of spin-orbit effects
induces mixing of the QMO states, as was shown in Ref.~\cite{Foyevtsova2013}.
Strong Coulomb repulsion can destroy the QMOs in favor of a description in terms of localized
states, however, some features of the QMOs may still be detected as it has been shown in
the analysis of optical conductivity of {\NIO} and $\alpha$-Li$_2$IrO$_3$~\cite{Li2015}.
%parity of the QMOs can be still probed
%and could both be dominatly described by QMOs and $j_{\rm eff}$ states. 
%Na2IrO3 lies in between a fully localized and fully itinerant description. 
Actually,  a description in terms of fully localized ($j_{\rm eff}$) or itinerant QMO scenarios strongly depends on 
a competition between the various energy scales involved (kinetic energy, crystal field splittings, spin-orbit coupling, and Coulomb repulsion).
%the energy scale considered: while the
%low-energy magnetic dynamics is compatible with a picture of localized moments characterized by %bond-dependent interactions, QMOs are the effective building blocks of the
%physics at binding energies larger than around 1 eV~\cite{Nembrini2016}.

For stoichiometric \ALIO~ our results show that the Ir-O hybridization 
 is moderate and a localized relativistic $j_{\rm eff} = 1/2$ magnetic model
is still valid for the description of the system. We also find that assuming the presence of Ag vacancies 
has an important impact on the experimental RIXS spectra due to the modification of Ir filling and hopping parameters.
We also show
that the extracted exchange parameters for the stoichiometric systems reproduce the experimentally observed spin spiral order. 

The paper is organized as follows. In Sec.~\ref{sec:dft} we discuss the electronic properties of various structures of \ALIO~ from the perspective of DFT calculations. In Sec.~\ref{sec:rixs} we calculate the RIXS spectra with the help of exact diagonalization of the multiorbital Hubbard model on finite clusters. In Sec.~\ref{sec:mag} the magnetic interactions are estimated and the magnetic properties are analyzed. Finally, in Sec.~\ref{sec:sum} we discuss and summarize our findings.  

\section{Density functional theory calculations}\label{sec:dft}

\begin{table*}[t]
\caption {Lattice parameters ($C2/m$), nearest-neighbor Ir-Ir distances (\AA), Ir-O distances (\AA) Ir-O-Ir bond angles ($^{\circ}$), crystal-field splittings (meV), and nearest-neighbor hopping integrals (meV) for the five structures. The $t_{2g}$ crystal fields $\Delta_1$, $\Delta_2$ denote, respectively, the on-site hopping between $d_{xz}$ and $d_{yz}$ orbitals, $d_{xy}$ and $d_{yz/xz}$ orbitals. $\Delta_3$ is the on-site energy of $d_{xy}$ minus $d_{yz/xz}$~\cite{Winter2016}. The labels $t_1$, $t_2$, $t_3$, and $t_4$ are given in Ref.~\cite{Rau2014, Winter2016} and shown in Fig.~\ref{fig:hoporbital}. The notations $t_{1\|}$, $t_{1O}$, $t_{1\sigma}$, and $t_{1\perp}$ are the same as in Refs.~\cite{Foyevtsova2013}.}
\centering\def\arraystretch{1.1}
\label{tab:hop}
\begin{ruledtabular}
\begin{tabular}{lrrrrrrrr}
Structure                 &\multicolumn{2}{c}{$S_1$}           & \multicolumn{2}{c}{$S_2$}     & \multicolumn{2}{c}{$S_3$}  & $S_4$  & $S_5$           \\
%Space group: &\multicolumn{8}{c}{$C2/m$}\\
a, b, c      &\multicolumn{4}{c}{5.283, 9.136, 6.486}           &\multicolumn{4}{c}{5.345, 9.014, 6.469} \\
$\alpha$, $\beta$, $\gamma$ & \multicolumn{4}{c}{90, 74.29, 90}           &\multicolumn{4}{c}{90, 105.42, 90} \\
\hline
Bond          & $Z$                & $X(Y)$            & $Z$      & $X(Y)$                 & $Z$          & $X(Y)$                & $Z$     & $Z$          \\
Ir-Ir          & 3.09             & 3.03            & 3.0479   & 3.0474               & 2.99       & 3.07                 & 2.99      & 2.99           \\
Ir-O        & 2.06            & 2.04                & 2.016  & 2.016              & 2.06       & 2.00                 & 2.06      & 2.06   \\
           &                & 2.02                &    & 2.018              &        & 2.16    &              &  \\
Ir-O-Ir        & 97.4             & 96.3            & 98.2   & 98.1                 & 93.3       & 94.8                & 93.3    &  93.3            \\
\hline                                                                                    
$\Delta_1$           & \multicolumn{2}{c}{-72.1}          & \multicolumn{2}{c}{-78.4}     & \multicolumn{2}{c}{-56.9}   & -17.1    &   12.7             \\
$\Delta_2$           & \multicolumn{2}{c}{-54.8}          & \multicolumn{2}{c}{-62.6}     & \multicolumn{2}{c}{-78.3}   & -23.0      & 2.0               \\
$\Delta_3$           & \multicolumn{2}{c}{55.7}           & \multicolumn{2}{c}{11.6}      & \multicolumn{2}{c}{157.0}    & 205.1     & 289.8              \\
\hline
Bond                 & $Z$                & $X(Y)$            & $Z$      & $X(Y)$                 & $Z$          & $X(Y)$          & $Z$     & $Z$         \\
$t_1$ ($t_{1\|})$    & -13.2            & -17.9           & -10.0  & -18.4                & 25.0      & -9.4           &  28.8    &     17.8       \\
$t_1^{\prime}$       &                 & 22.4            &       & 9.4                  &               &  32.6  &     &             \\
$t_2$ ($t_{1O}$)     & 168.6            &154.2            & 185.0  &189.4                 & 203.4     & 109.2          &  310.1   & 281.9    \\
$t_3$ ($t_{1\sigma}$)& -97.0            &-158.2           & -75.5  &-87.1                 & -242.8    & -158.1         &  -137.2  & -124.7        \\
$t_4$ ($t_{1\perp}$) & -27.7            &-2.3             & -39.7  &-23.7                 & -48.1     & -34.4          &  -6.7    & -9.4           \\
$t_4^{\prime}$       &                 &7.7              &       &-18.4                 &             & -26.9  &  -5.9    &     -9.3      \\
\end{tabular}
\end{ruledtabular}
\end{table*}

\begin{figure}[htpb]
\includegraphics[angle=0,width=\linewidth]{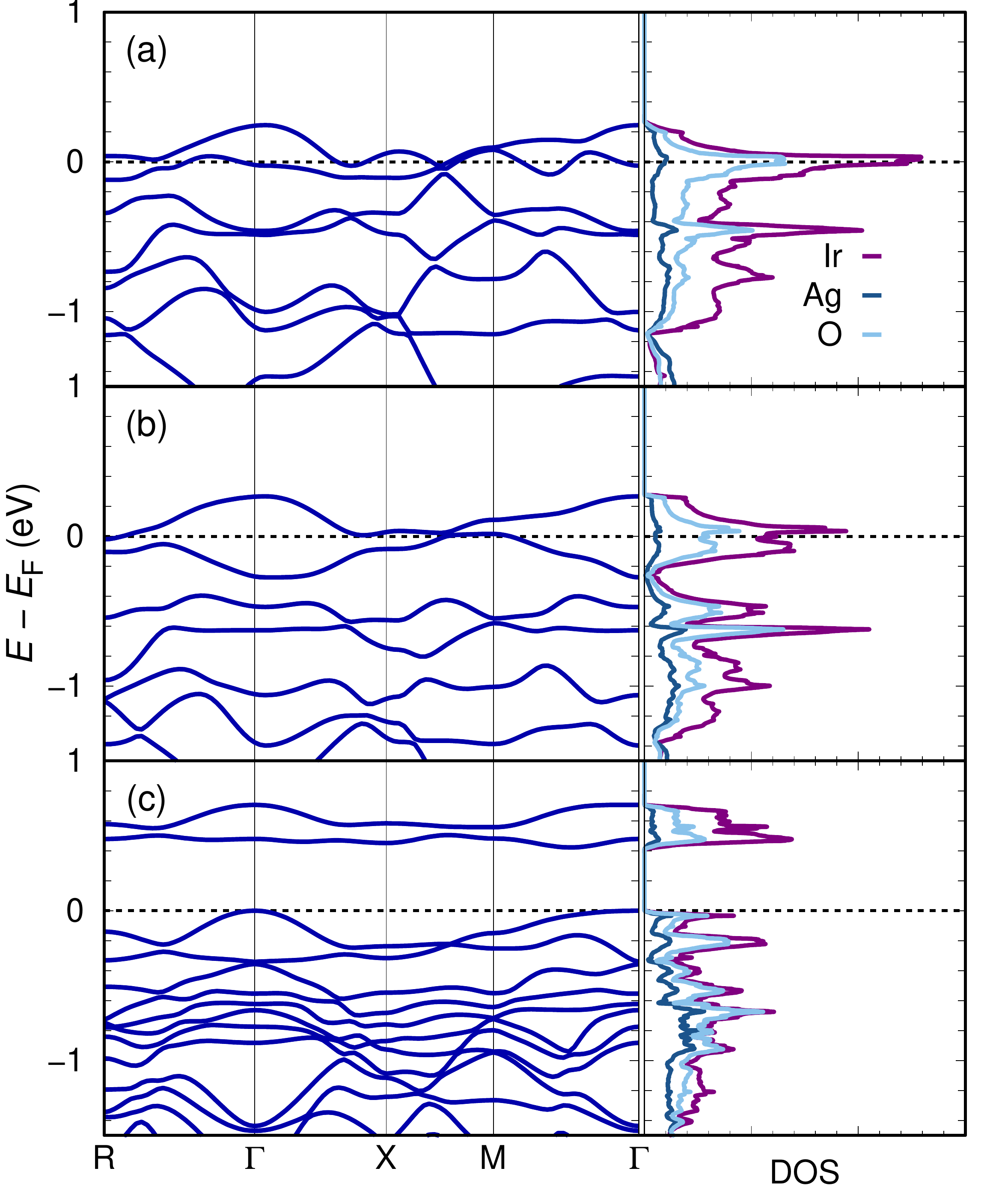}
\caption{Band structure and partial density of states for the relaxed structure $S_2$ within (a) GGA, (b) GGA+SO, and (c) GGA+SO+$U$, respectively,
obtained with the LAPW basis~\cite{Wien2k, Blaha2020}.
In the GGA+SO+$U$ calculation we considered a stripy magnetization as shown
in Fig.~\ref{fig:structure} with doubling of the unit cell.}
\label{fig:banddosrelax}
\end{figure}

The crystal structure of \ALIO~is displayed in Figs.~\ref{fig:structure} (a) and~\ref{fig:structure} (b). 
Edge sharing IrO$_6$ octahedra build the hexagonal planes in \ALIO~
with Li atoms in the hexagonal center while Ag  atoms are placed between the layers. Since the structural details are different in various experiments, we consider here various structures of \ALIO~: (i) the experimental structure from Ref.~\cite{Bahrami2021} ($S_1$); (ii) the
corresponding relaxed structure within DFT ($S_2$); (iii) the experimental structure from Ref.~\cite{Torre2021} ($S_3$);
and
(iv) to investigate the effects of possible charge disbalance through Ag vacancies, we remove one Ag along the $Z$ bond from the $S_3$ structure resulting in structure $S_4$
with one Ag vacancy per six Ag atoms in the 
unit cell as shown in Fig.~\ref{fig:structure} (c), and we consider as well the case of one Ag vacancy per two Ag atoms per unit cell
as shown in 
Fig.~\ref{fig:structure} (d) for the $Z$ bond, that we denote structure $S_5$.
Starting from the $S_3$ structure, the bond lengths, and angles of the resultant relaxed structure 
have values  between those of  $S_2$ and $S_3$ and its total energy is higher than that of $S_2$.
We therefore consider in what follows only $S_2$ as the theoretical relaxed structure.

We note that the vacancy cases considered correspond to
 a much larger concentration of vacancies than
the possible concentrations present in the material. Simulation of smaller concentrations would require large supercell calculations. Since we are interested in evaluating the possible effects emerging
from the local presence of Ag vacancies around Ir,
we consider here the very extreme cases $S_4$ and $S_5$ where
not only the Ir filling of the Ag neighboring Ir is affected, but the Ir-O hoppings as well.
In $S_4$ [Fig.~\ref{fig:structure} (c)] the Ag vacancies induce 
on the $Z$ bond a filling Ir$^{4.5^+}$ (occupation $d^{4.5}$). In $S_5$ [Fig.~\ref{fig:structure} (d)] the
Ir filling on the $Z$ bond is Ir$^{5^+}$ (occupation $d^{4}$). All the structures are in
the $C2/m$ symmetry as shown in Fig.~\ref{fig:structure}.
The corresponding lattice parameters, bond lengths, and Ir-O-Ir angles 
for the five structures are displayed in Table.~\ref{tab:hop}. The anisotropy between the $X$ ($Y$) and $Z$ bonds are found to be strong in the two experimental structures $S_1$ and $S_3$ while weak in the relaxed structure $S_2$. For the $S_3$ structure, the bond ratio $l_l$/$l_s$ $\sim$ 1.03, where $l_l$ denotes the length of the long bond (X and Y) and $l_s$ the length of the short bond ($Z$) is smaller than the strong dimerization case under pressure in \rucl~\cite{Biesner2018} with $l_l$/$l_s$ $\sim$ 1.25.  

The $S_2$ structure is obtained from the experimental structure $S_1$ by fixing the lattice parameters and relaxing the atomic coordinates using the Vienna {\it ab initio} simulation package (VASP)~\cite{Kresse1996,Hafner2008}. To keep consistent with previous calculations~\cite{Li2015} for {\LIO}, we considered relativistic effects as well as
contributions of the Coulomb repulsion~\cite{Dudarev1998} ($U_{\rm eff}$ = 2.4 eV) within GGA+SO+$U$. We adopted a cutoff energy of 520 eV and Monkhorst-pack $k$-points generated with 8 $\times$ 6 $\times$ 8. The choice of $U_{\rm eff}$ = 2.4 eV was done
following Ref.~\cite{Li2015}, which was determined by the gap of the electronic structures in \NIO. We also considered $U_{\rm eff} = 2$ eV and 3 eV, but the resultant relaxed structures
are not significantly affected by the values. 

The band structures were obtained from full-potential linearized augmented plane-wave (LAPW) calculations~\cite{Wien2k}. We chose the basis-size controlling parameter RK$_{\rm max} = 8$ and a mesh of 500 {\bf k} points in the first Brillouin zone (FBZ) of the primitive unit cell. The density of states (DOS) were computed with 1000 {\bf k} points in the full Brillouin zone. 

We start with the band structures and partial DOS 
within GGA, GGA+SO, and GGA+SO+$U$ obtained from LAPW calculations for the relaxed structure $S_2$ as presented in Figs.~\ref{fig:banddosrelax} (a) to ~\ref{fig:banddosrelax} (c). 
The DOS around the Fermi level in GGA is dominated by Ir $t_{2g}$ and O states
and it includes contributions from Ag, in contrast to {\LIO} and {\HLIO} where the DOS around the Fermi level has only Ir and O contributions.
From this we expect that Ag will affect the oxygen assisted 
Ir-Ir hoppings stronger than Li. 

The inclusion of $U$ within the GGA+SO+$U$ approach in the stripy magnetic configuration  [see Fig.~\ref{fig:structure} (b)], which has the lowest energy within all collinear magnetic configurations, opens a gap of 409 meV [Fig.~\ref{fig:banddosrelax} (c)]. 
We note that the electronic properties for the two experimental structures $S_1$ and $S_3$ are similar
to those obtained for the $S_2$ structure with Ag contribution around the Fermi level and insulating behavior within GGA+SO+$U$.

Valuable information on the hybridization patterns can be obtained
from the analysis of the hopping parameters extracted from GGA. Table~\ref{tab:hop} displays the
 hopping parameters between the Ir $5d$ $t_{2g}$ orbitals computed  via the Wannier function projection method~\cite{Foyevtsova2013, Winter2016} for the five structures. In terms of the $t_{2g}$ $d$-orbital basis:
\begin{align}
\vec{\mathbf{c}}_i^\dagger = \left(c^\dagger_{i,yz,\uparrow} \  c^\dagger_{i,yz,\downarrow} \ c^\dagger_{i,xz,\uparrow} \  c^\dagger_{i,xz,\downarrow} \ c^\dagger_{i,xy,\uparrow} \  c^\dagger_{i,xy,\downarrow}\right),
\end{align}
where $c_{i,a}^\dagger$ creates a hole in orbital $a\in\{d_{yz},d_{xz},d_{xy}\}$ at site $i$, the crystal field terms can be written as:
\begin{align}
\mathcal{H}_{\rm CF}= - \sum_i \vec{\mathbf{c}}_{i}^\dagger\left\{\mathbf{E}_i\otimes \mathbb{I}_{2\times 2}\right\}\vec{\mathbf{c}}_i,
\end{align}
where $\mathbb{I}_{2\times 2}$ is the $2 \times 2$ identity matrix (for the spin variables); the crystal field tensor $\mathbf{E}_i$ is constrained by local two-fold symmetry at each Ir site to be:
\begin{align}
\mathbf{E}_i = \left(\begin{array}{ccc} 0&\Delta_{1}&\Delta_{2} \\ \Delta_{1}&0&\Delta_{2} \\ \Delta_{2} & \Delta_{2} & \Delta_{3}\end{array} \right)
\end{align}
The $t_{2g}$ crystal fields
$\Delta_1$, $\Delta_2$ denote the on-site hopping between $d_{xz}$ and
$d_{yz}$ orbitals, and between $d_{xy}$ and $d_{yz/xz}$ orbitals,
respectively (Table~\ref{tab:hop}). $\Delta_3$ is the on-site energy
of $d_{xy}$ minus that of $d_{yz/xz}$~\cite{Winter2016}. 
There are large trigonal distortions ${\Delta}_1$ and ${\Delta}_2$ due to Ag atoms, which induce an anisotropic crystal field on Ir by distorting the Ir-O octahedra. Replacing Ag by Li and keeping the local geometry reduces the values to -49 meV ($\Delta_1$) and -41 meV ($\Delta_2$) for the $S_2$ structure, which is closer to the  {\LIO} results~\cite{Winter2016}. We observe that the tetragonal distortion ${\Delta}_3$ is 157 meV for 
the $S_3$ structure, which is much larger than in the case of $S_1$, $S_2$ and {\LIO} (-5.5 meV)~\cite{Winter2016} 
and closer to that of $\gamma$-Li$_2$IrO$_3$~\cite{Li2017}. In this case the $t_{2g}$ crystal field is of the same order of magnitude as the spin-orbit coupling $\lambda$ and this has significant effects on the local magnetic interactions
as we will show further below. In the structure $S_4$ with Ag vacancies,
on the $Z$-bond the crystal field  ${\Delta}_3$ is enhanced while $\Delta_1$ and $\Delta_2$ are somewhat suppressed in comparison to the $S_3$ structure.

\begin{figure}[tpb]
\includegraphics[angle=0,width=0.9\linewidth]{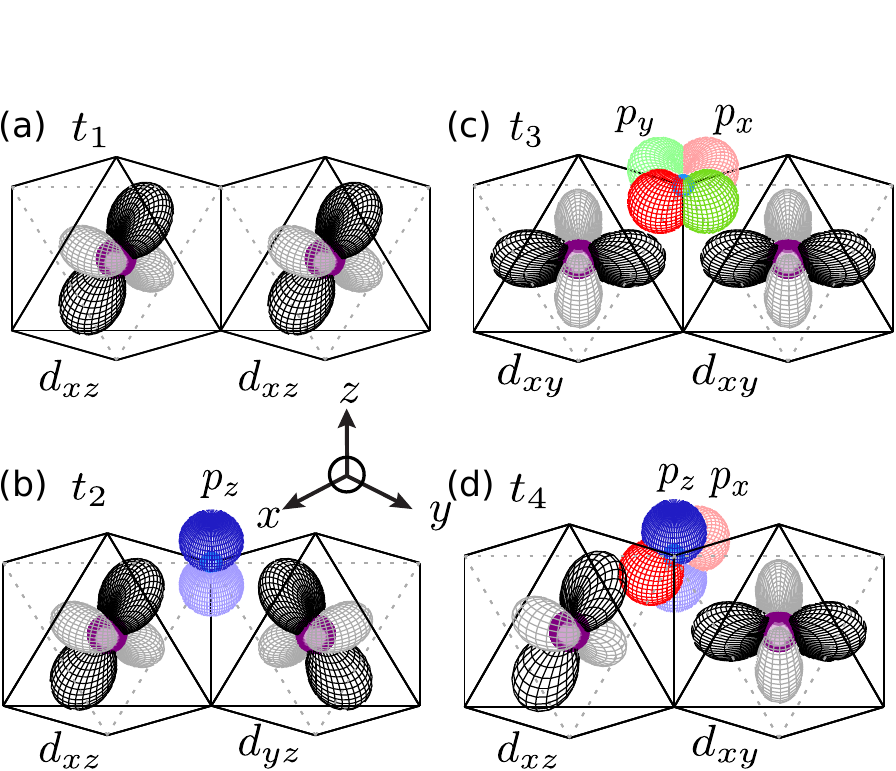}
\caption{Geometry of nearest-neighbor hopping integrals (a) $t_1$, (b) $t_2$, (c) $t_3$, and (d) $t_4$ for the $Z$ bond~\cite{Li2018}.
Both $t_2$ and $t_3$ include contributions of
 oxygen assisted hopping and direct hopping. $t_2$ is dominated by oxygen assisted hopping while $t_3$ is dominated by direct hopping.
In \ALIO\ the oxygen assisted part of $t_3$ is strongly affected by Ag-O hybridization.}
\label{fig:hoporbital}
\end{figure}

The nearest neighbor hopping parameters $t_1$, $t_2$, $t_3$ and $t_4$ are defined in Refs.~\cite{Rau2014, Winter2016} and shown in Fig~\ref{fig:hoporbital}
for the $Z$-bond in terms of $t_1 = t_{xz,xz} = t_{yz,yz}$, $t_2 = t_{xz,yz} = t_{yz,xz}$, $t_3 = t_{xy,xy}$, and
$t_4 = t_{xz,xy} = t_{yz,xy} = t_{xy,xz} = t_{xy,yz}$. 
In terms of the $t_{2g}$ $d$-orbital basis, the hopping Hamiltonian is most generally written as
\begin{align}
\mathcal{H}_{\rm hop}=  - \sum_{ij} \vec{\mathbf{c}}_{i}^\dagger \ \left\{\mathbf{T}_{ij} \otimes \mathbb{I}_{2\times 2}\right\}\ \vec{\mathbf{c}}_j,
\end{align}
with the hopping matrices $\mathbf{T}_{ij}$ defined for each bond connecting sites $i,j$. The hopping integrals for the nearest neighbor $Z$-bond ($C_{2h}$ symmetry) are written as~\cite{Winter2016}
\begin{align}
\mathbf{T}_{Z} = \left(\begin{array}{ccc} t_1 & t_2 & t_4 \\ t_2 & t_1 & t_4 \\ t_4 & t_4 & t_3 \end{array} \right)
\end{align}
where $X$- and $Y$-bonds are of lower symmetry ($C_i$), and therefore $t_1$ ($t_4$) split into two values which are labeled as $t_1$ and $t_1^{\prime}$ ($t_4$ and $t_4^{\prime}$). The hopping matrices become:
\begin{align}
\mathbf{T}_{X} = \left(\begin{array}{ccc} t_3 & t_4^{\prime} & t_4 \\ t_4^{\prime} & t_1^{\prime} & t_2 \\ t_4 & t_2 & t_1 \end{array} \right),
\mathbf{T}_{Y} = \left(\begin{array}{ccc} t_1^{\prime} & t_4^{\prime} & t_2 \\ t_4^{\prime} & t_3 & t_4 \\ t_2 & t_4 & t_1 \end{array} \right).
\end{align}

For the experimental structure $S_1$, there is a strong anisotropy between $Z$ and $X/Y$-bonds. For the $Z$-bond, $t_2$ $>$ $|t_3|$ while for the $X$ bond, $|t_3|$ is slightly larger than $t_2$.  These anisotropies
are even more pronounced for the experimental structure $S_3$. The magnitude of the direct hopping $|t_3|$ is much larger than the oxygen assisted one $t_2$. After structural relaxation, the bond anisotropy is suppressed in $S_2$. The hopping parameters, which are very sensitive to the structural details, are slightly different between the $Z$ bond and $X/Y$-bonds. The ratio of $t_2/|t_3|$ mainly depends on the angle of Ir-O-Ir~\cite{Winter2016}. Since $S_2$ has the largest Ir-O-Ir angle within the three structures, it hence has the largest $t_2/|t_3|$. The Ir-O-Ir angle in the $S_2$ structure is 98$^{\circ}$, close to the angle of {\NIO}. Ir-O bond lengths are 
2.0 $\sim$ 2.18 \AA~for $S_1$-$S_3$ in {\ALIO}   and 2.06 $\sim$ 2.08 \AA~in {\NIO} . However, $t_2$ is much smaller in {\ALIO} than in {\NIO} (264 meV) while the magnitude of $|t_3|$ is much larger than in {\NIO} (26.6 meV). Therefore the effect of Ag is to enhance the Ag-O hybridization and, correspondingly, the oxygen mediated $d$-$d$ hopping integrals involving single (multiple) O $p$-orbitals are suppressed (enhanced). Taking the $Z$-bond as an example shown in Fig.~\ref{fig:hoporbital}, $t_2$ is reduced due to suppression of the hopping paths like
Ir($d_{xz}$)$ \to $O($p_z$)$ \to$Ir($ d_{yz}$). Similarly, $|t_3|$ and $|t_4|$ are enhanced through hopping paths such as Ir($d_{xz}$)$ \to
$O($p_z$)$\to$Ag($s$)$\to$O($p_x$)$\to$Ir($ d_{xy}$). 

This is opposite when vacancies are introduced. 
The Ag vacancy in
the $S_4$ structure enhances $t_2$ in the $Z$ bonds from 203.4 meV to 310.1 meV and reduces $|t_3|$ from 242.8 meV to 137.2 meV. This is because Ir and O have stronger hybridization without Ag. For the $X$ ($Y$) bond (not shown), the symmetry of the two Ir-O-Ir hopping paths is broken by the asymmetric Ag atom positions, leading to different $t_2$. 
\begin{figure}[t]
\includegraphics[angle=0,width=\linewidth]{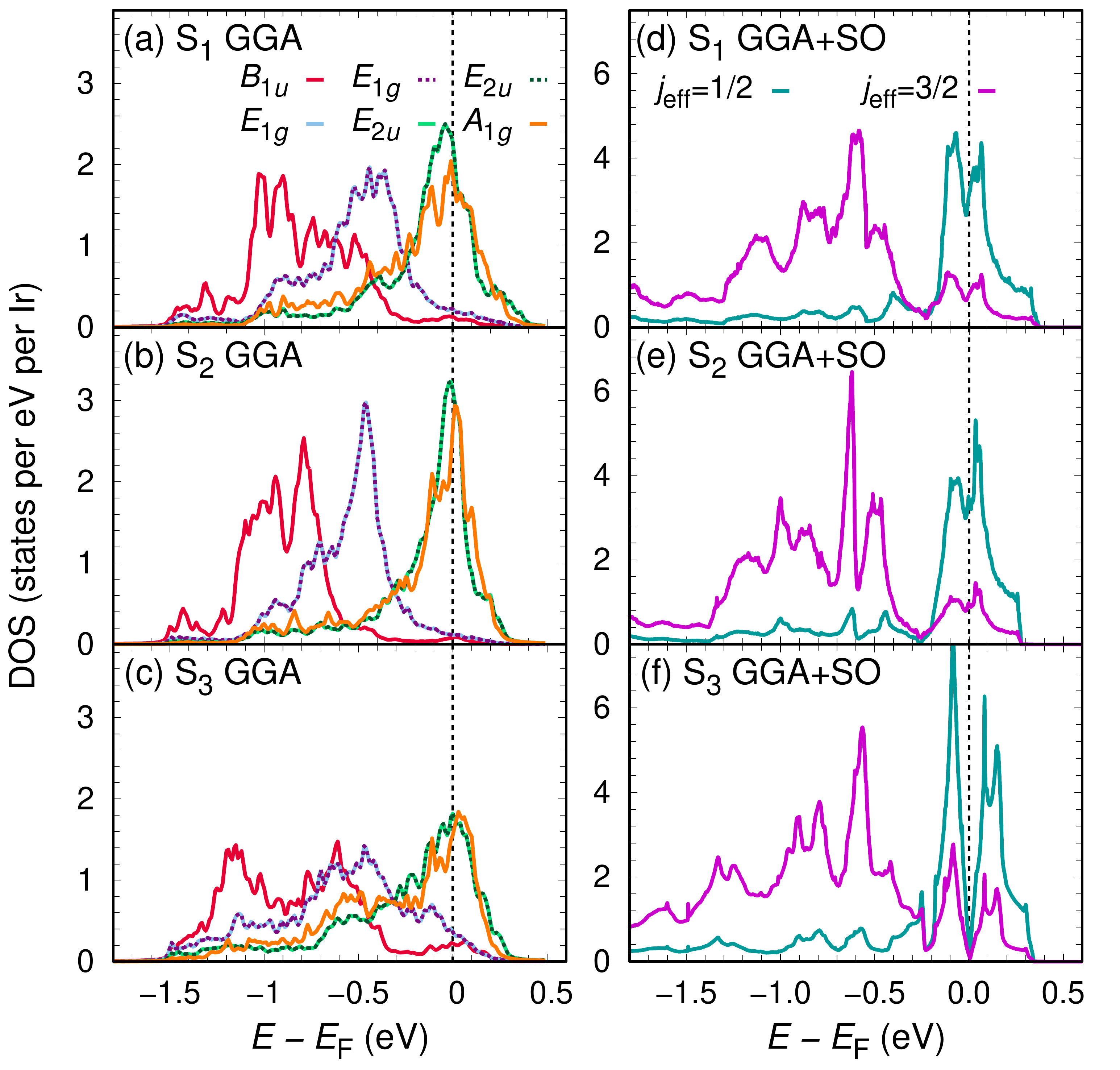}
\caption{(a)-(c) Nonrelativistic GGA density of states projected onto quasimolecular orbitals and (d)-(f) relativistic GGA+SO density of states projected onto the relativistic $j_{\rm eff}$ basis for structures $S_1$ - $S_3$.}
\label{fig:qmoreldos}
\end{figure}

To discuss the choice of basis for the effective spin Hamiltonian, we first display in Figs.~\ref{fig:qmoreldos} (a) to~\ref{fig:qmoreldos} (c) the nonrelativistic density of states within GGA projected onto the quasimolecular-orbital (QMO)~\cite{Mazin2012, Foyevtsova2013} basis for the structures $S_1$ - $S_3$. We observe that the separation of the density of states into isolated narrow bands of unique QMO characters is much less evident in $S_1$-$S_3$ than in {\NIO}~\cite{Mazin2012} and resembles the case of \LIO~\cite{Li2015}. 
For instance, for 
the $S_3$ structure we observe a strong mixing of QMO states due to a smaller $t_2/|t_3|$
than in {\NIO}. 
When spin-orbit effects are included, 
while  for {\NIO} (with $t_2 \sim $ 0.27 eV, $|t_3| \sim $ 0.025 eV~\cite{Foyevtsova2013})
both, the relativistic basis and QMO provide
a good description of the electronic properties, 
in \ALIO~$t_2$ is reduced to 0.15 $\sim$ 0.2 eV and $|t_3|$ increases, resulting in the suppression of QMO and enhancement of the relativistic basis. We therefore present in Figs.~\ref{fig:qmoreldos} (d) to~\ref{fig:qmoreldos} (f) the electronic structure within GGA+SO projected to the relativistic $j_{\rm eff}$ basis. The spin-orbit coupling largely destroys the QMOs and leads instead to the formation of relativistic $j_{\rm eff} = 1/2$ orbitals with a small contribution from $j_{\rm eff} = 3/2$ around the Fermi level. 

\section{RIXS spectra}\label{sec:rixs}

\begin{figure}[htpb]
\includegraphics[angle=0,width=\linewidth]{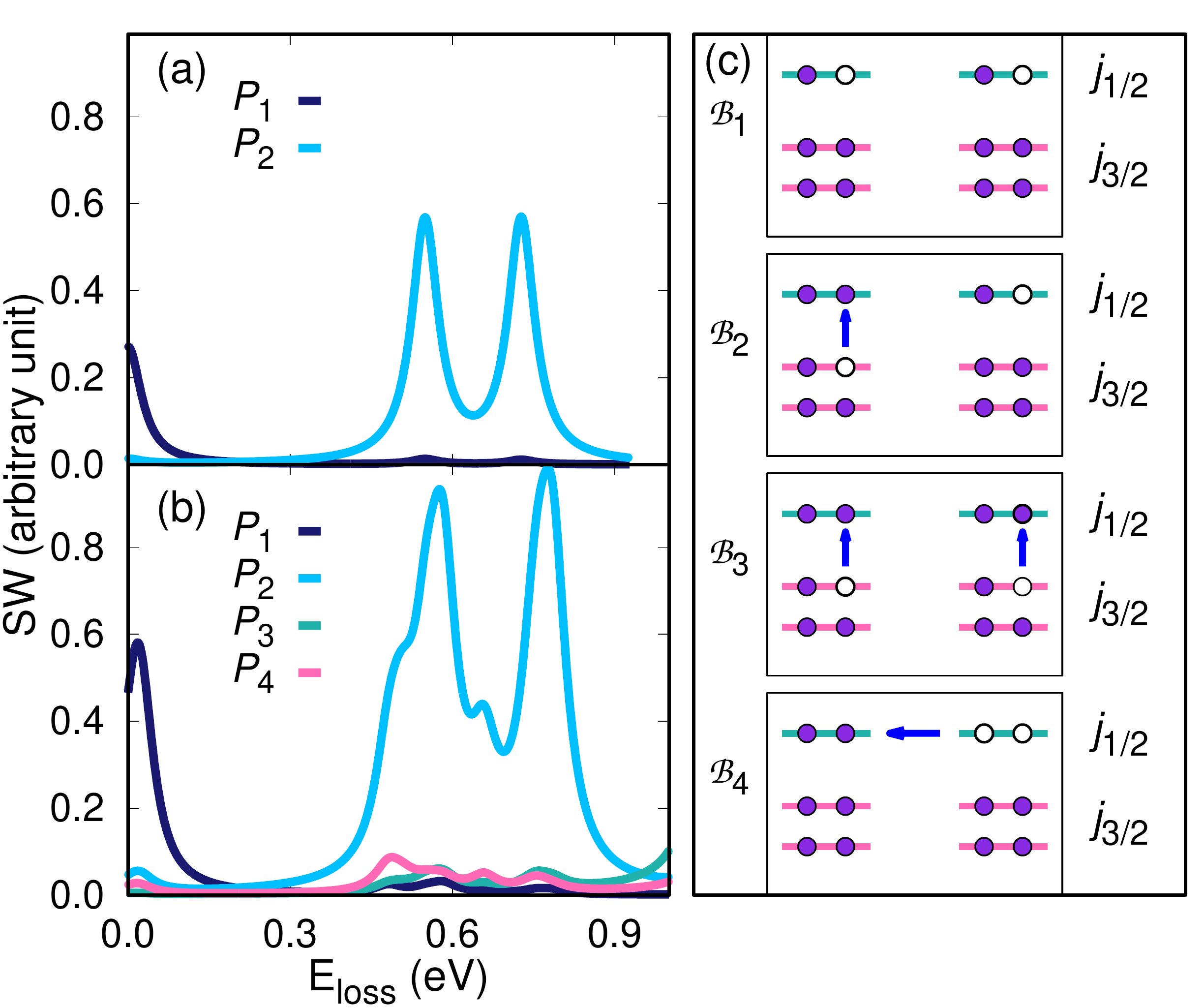}
\caption{Spectral weight of various states [Eq.~\ref{eq:pes}] obtained from performing (a) one-site and (b) two-site cluster calculations  for the experimental structure $S_3$. $P_1$ indicates the ground state, $P_2$ is a local exciton, $P_3$ are multiple excitons while $P_4$ are all the projections including $d^4$-$d^6$. (c) Schematic diagrams of the lowest-energy subspaces $\mathcal{B}_1$, $\mathcal{B}_2$, $\mathcal{B}_3$  and $\mathcal{B}_4$ as defined in the main text.}
\label{fig:subspace}
\end{figure}

Based on the density of states from DFT, we calculated
the joint density of states (see the Appendix~\ref{app}), which corresponds to the peak positions of the RIXS spectra. However, as discussed in Ref.~\onlinecite{Li2017}, DFT does not fully capture the effects originating from correlations beyond GGA+SO+$U$, which are expected to be relevant when analyzing electronic excitations.
Specifically, DFT does not correctly capture the spin-multiplicity associated with the localized states. 
To compare to the experimental RIXS spectra, we perform exact diagonalization
of one-site and two-sites clusters of the Ir $t_{2g}$-only Hamiltonian

\begin{eqnarray}
\mathcal{H}_{\rm tot} = \mathcal{H}_{\rm hop}+\mathcal{H}_{\rm CF}+\mathcal{H}_{\rm SO} +
\mathcal{H}_{U}
\end{eqnarray}
consisting of the kinetic hopping term $\mathcal{H}_{\rm hop}$, the
crystal field splitting $\mathcal{H}_{\rm CF}$, spin-orbit coupling
$\mathcal{H}_{\rm SO}$, and Coulomb interaction $\mathcal{H}_{U}$ contributions. In terms of the $t_{2g}$ basis introduced above, the spin-orbit coupling (SO) is described by:
\begin{align}
\mathcal{H}_{\rm SO}=\frac{\lambda}{2} \sum_i \vec{\mathbf{c}}_{i}^\dagger\left(\begin{array}{ccc} 0 & -i \sigma_z & i \sigma_y \\ i \sigma_z & 0 & -i\sigma_x \\ -i \sigma_y & i\sigma_x & 0\end{array} \right)\vec{\mathbf{c}}_i
\end{align}
where $\sigma_\mu$, $\mu=\{x,y,z\}$ are Pauli matrices. The Coulomb terms are:
\begin{align}
\mathcal{H}_{U}& \ = U \sum_{i,a} n_{i,a,\uparrow}n_{i,a,\downarrow} + (U^\prime - J_{\rm H})\sum_{i,a< b, \sigma}n_{i,a,\sigma}n_{i,b,\sigma} \nonumber \\
&+ U^\prime\sum_{i,a\neq b}n_{i,a,\uparrow}n_{i,b,\downarrow} - J_{\rm H} \sum_{i,a\neq b} c_{i,a\uparrow}^\dagger c_{i,a\downarrow} c_{i,b\downarrow}^\dagger c_{i,b\uparrow}\nonumber \\ & + J_{\rm H} \sum_{i,a\neq b}c_{i,a\uparrow}^\dagger c_{i,a\downarrow}^\dagger c_{i,b\downarrow}c_{i,b\uparrow} 
\end{align}
where $J_{\rm H}$ gives the strength of Hund's coupling, $U$ is the {\it intra}orbital Coulomb repulsion, and $U^\prime=U-2J_{\rm H}$ is the {\it inter}orbital repulsion. For $5d$ Ir we take $U=1.7$ eV, $J_{\rm H}=0.3$ eV\cite{Yamaji2014, Winter2016}. 
Based on the eigenenergies, we analyzed the states. In one-site and two-site clusters, we consider states with a total of one hole or two holes in the $t_{2g}$ orbitals, respectively. Each Ir site contains six relativistic orbitals consisting of two $j_{\rm eff}=1/2$ and four $j_{\rm eff}=3/2$ levels. As in Refs.~\cite{BHKim2014, Li2017}, the many-body basis states for the cluster can be divided into several subspaces $\mathcal{B}_i~(i = 1, 2, 3, 4)$ based on the occupancy of the various orbitals and sites as shown in Fig.~\ref{fig:subspace} (c).
The subspace $\mathcal{B}_1$ contains all states with $(j_{3/2})^4(j_{1/2})^1$ occupancy at every site, which represent a significant contribution to the ground state and low-lying magnon-like spin excitations. From these configurations, the promotion of a single electron via {\it on-site} $j_{3/2}\rightarrow j_{1/2}$ generates subspace $\mathcal{B}_2$, containing all states with a single spin-orbital exciton; the characteristic
excitation energy for such states is given by $\Delta E_2 \sim 3\lambda/2 \sim 0.6$ eV if the crystal-field and hopping parameters are zero. For the two-site cluster, the states with two excitons are grouped into subspace $\mathcal{B}_3$ with energies $\Delta E_3 \sim 2 \Delta E_2$, and the basis states with site occupancy 
$d^4$-$d^6$ belong to $\mathcal{B}_4$. We project the exact cluster eigenstates $\phi_m$ on different subspaces $\mathcal{B}_i$:
\begin{align}
\Gamma^m_i = \sum_{b \in \mathbf{\mathcal{B}_i}} \left | \left\langle \phi_m | b \right\rangle \right |^2, %\left |
\label{eq:pescof}
\end{align}
and  take the spectral weight (SW) of the projected excitation spectra $P_i$~\cite{BHKim2014, Li2017}:
\begin{align}
P_i (E_{\rm loss}) =\sum_m \Gamma^m_i \delta \left(E_{\rm loss} - E_m\right),
\label{eq:pes}
\end{align}
where $E_{\rm loss}$ is the energy transfer from the ground state to all other states. $P_i$ ($i$ = 1, 2, 3, 4) are shown in Fig.~\ref{fig:subspace} (a) for one site and Fig.~\ref{fig:subspace} (b) for two sites ($Z$-bond) cluster calculations. For the one-site cluster, as expected, the ground state around 0 eV has dominant $\mathcal{B}_1$ character (large $P_1$), and the peaks centered at 0.55 and 0.73 eV have dominant $\mathcal{B}_2$ character. For two-site cluster of the $Z$-bond, in addition to the ground state around 0 eV, $P_1$ has a peak around 0.02 eV indicating low-lying magnon-like spin excitations. Regarding higher excitations, $\mathcal{B}_2$ is weakly mixed with the multi-particle $\mathcal{B}_3$ and $\mathcal{B}_4$ excitations via intersite hopping. $P_2$ has peaks of 0.58, 0.78 eV and an additional shoulder at 0.49 eV.  Similar results were obtained in Ref.~\cite{BHKim2014} in the analysis of the excitation spectra of \NIO. For the two-site cluster calculation of $X$-bond, the peaks are close to $Z$-bond, but the 0.49 eV one disappears due to smaller hopping integrals.

\begin{figure}[tpb]
\includegraphics[angle=0,width=\linewidth]{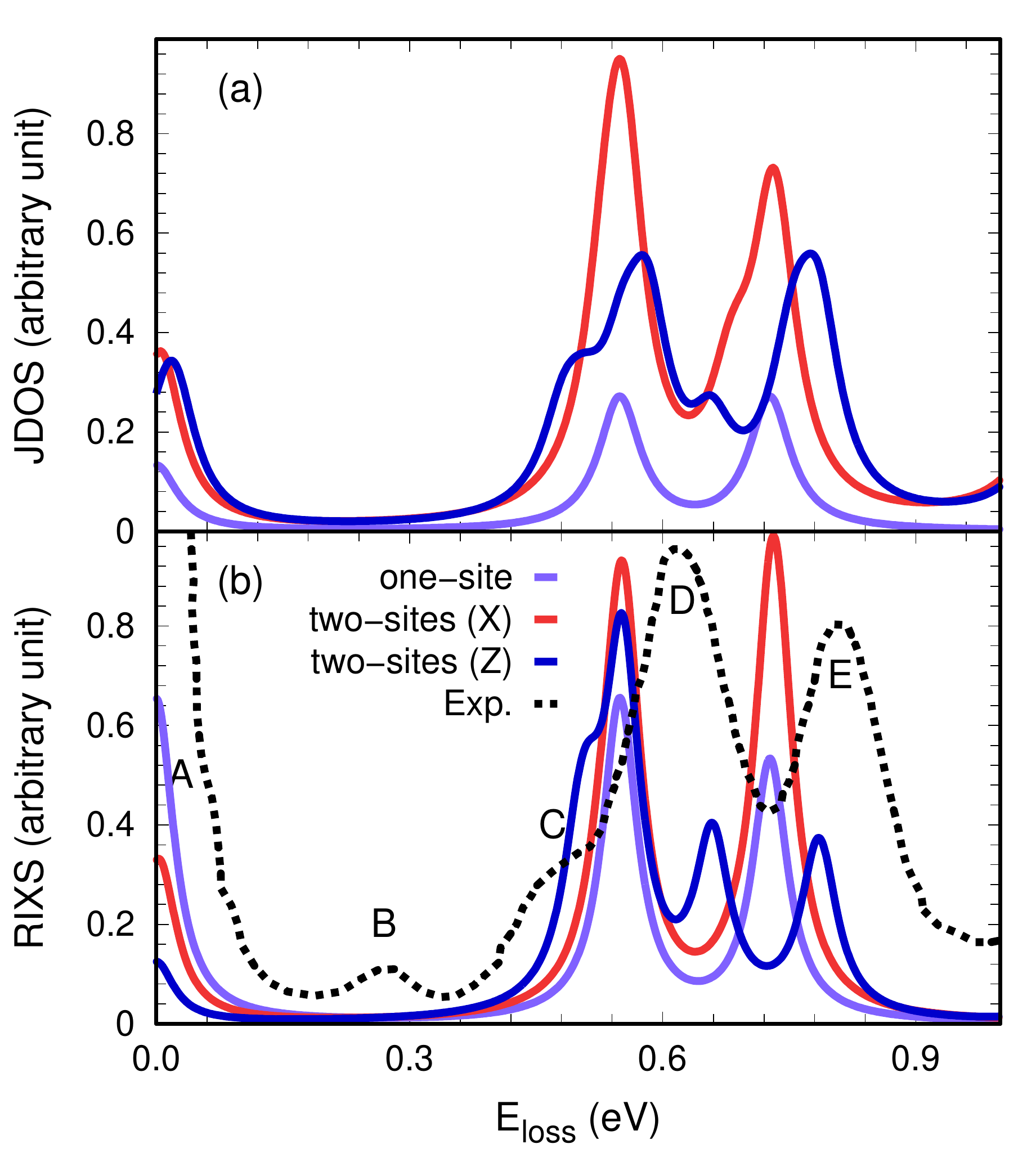}
\caption{(a) JDOS and (b) RIXS results obtained from one-site and two-site cluster calculations (see main text) for the experimental structure $S_3$. The experimental data
from Ref.~\cite{Torre2021} are shown
with a black dashed line where A, B, C, D, E indicate the experimental observed peaks. 
}
\label{fig:jdosrixs}
\end{figure}

The joint density of states (JDOS) is defined as
\begin{equation}\begin{split}
    {\rm JDOS}(E_{\rm loss}) \propto \delta (E_{\rm loss} - E_f + E_g)
\end{split}
\label{eq:jdos}
\end{equation}
where $E_g$ and $E_f$ are the ground states and excited states, respectively.
%In order to identify each peak, 
%The calculated JDOS 
The RIXS spectra are given by  
\begin{equation}\begin{split}
{\rm RIXS} (E_{\rm loss}) \propto \sum_{f} |\sum_i e^{-i \mathbf{Q} \cdot \mathbf{R_i}} \langle f|F_i \rangle|^2  \delta (E_{\rm loss} - E_f + E_g)
\end{split}
\label{eq:rixs}
\end{equation}
where the final state is generated by the RIXS process:

\begin{equation}\begin{split}
    |F_i\rangle = D_i^{\dagger} \frac{1}{E_{\rm in} - H_{\rm inter} + E_g + i\Gamma_c/2} D_i |g\rangle.
\end{split}
\end{equation}
Here $g$ ($f$) are the ground (excited) eigenstates of $H_{\rm tot}$, and $E_{\rm in}$ is the energy of the incident x-rays. $D_i$ is the dipolar transition operator from $2p$ to $5d$ shell on the Ir site, and $H_{inter}$ is the intermediate configuration of the RIXS process where a $2p$ core-hole is created and $\Gamma_c$ is the core-hole life time broadening. $\mathbf{Q}$ represents the wave vectors of the incident and outgoing photons and $\mathbf{R_i}$ are the positions of Ir sites. 
Here we use the EDRIXS software package~\cite{Wang2019} for the calculations.

In our calculation, we set 2 $\theta$ = 90$^{\circ}$ and fix the incident beam polarization to lie in the scattering plane and average over the outgoing direction in and perpendicular to the scattering plane.
JDOS and RIXS have the same peak positions while the weights are different. Including the matrix elements in RIXS, some JDOS peaks are enhanced while others are suppressed. The calculated JDOS and RIXS spectra obtained for the experimental structure $S_3$ are compared in Fig.~\ref{fig:jdosrixs} with the experimental measurements. 
%The experimental RIXS spectra are presented in Fig.~\ref{fig:jdosrixs} (b). 
There are five peaks from the experiment at A $\sim$ 0.029 eV, B $\sim$ 0.27 eV, C $\sim$ 0.47 eV, D $\sim$ 0.623 eV, and E $\sim$ 0.811 eV. The peaks D and E which indicate the local excitations from $j_{1/2}$ to $j_{3/2}$ could be obtained both by our one-site and two-site calculations. The peaks A corresponding to magnon-like spin excitations and C arising from the mixing of local exciton state $\mathcal{B}_2$ with other states could be captured by our two-site calculation. We observe that the peak B does not appear in the calculated results. In Ref.~\cite{Torre2021} a large $t_2$ $\sim$ 0.525 eV had to be assumed to reproduce the data. The JDOS and RIXS spectra were also calculated and compared for $S_1$ and $S_2$ (see the Appendix~\ref{app}) and find that the peaks C and D are robust for the three structures but E only appears in $S_3$.

%We are discussing this in more detail in what follows. 

\begin{figure}[htpb]
\includegraphics[angle=0,width=\linewidth]{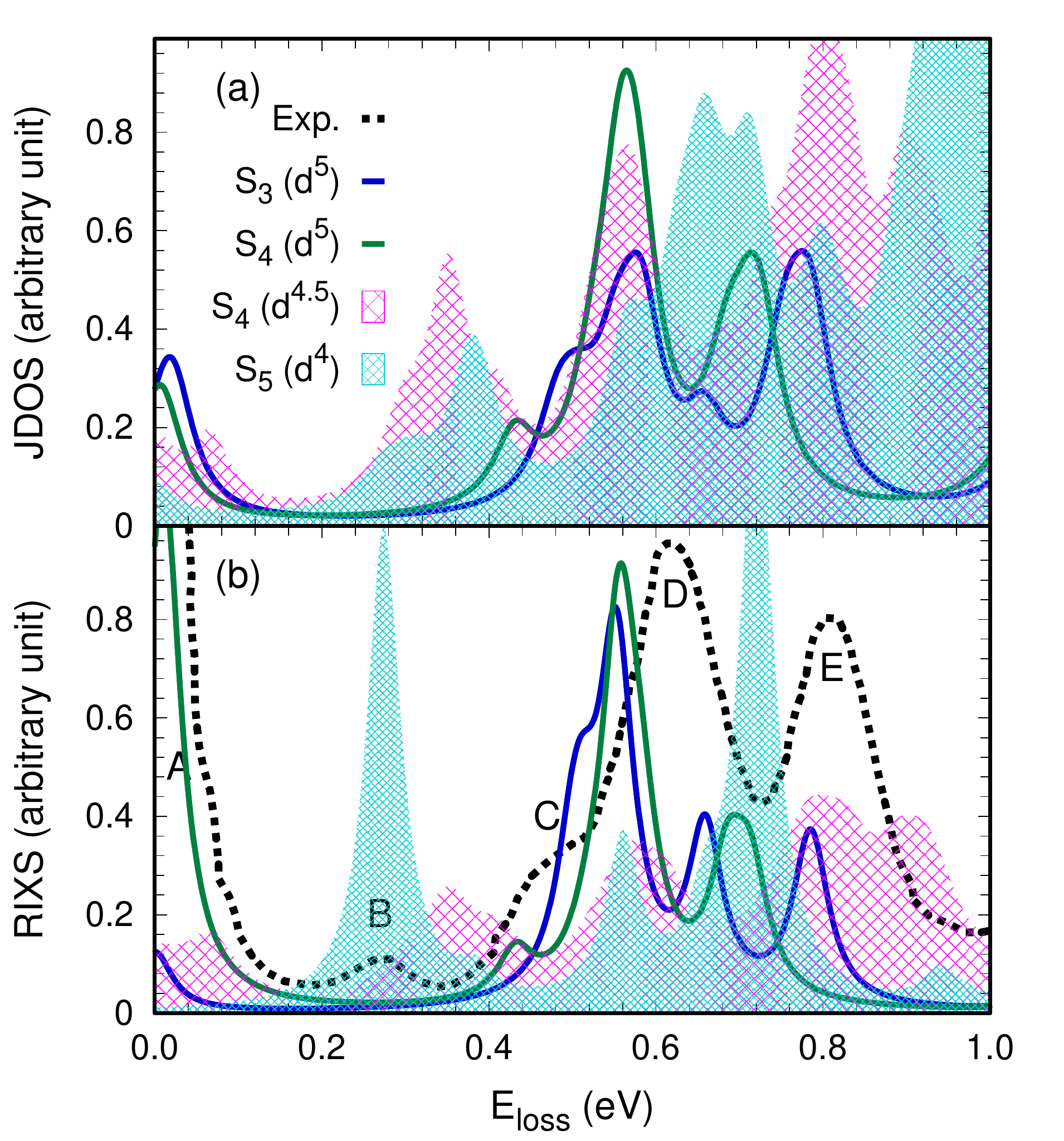}
\caption{(a) JDOS and (b) RIXS results obtained from two-site cluster ($Z$-bond)
calculations for (i) the structure $S_3$ with Ir occupancies $d^5$-$d^5$, 
(ii) the structure $S_4$ with Ir occupancies $d^5$-$d^5$,
(iii) the structure $S_4$ with Ir occupancies $d^{4.5}$-$d^{4.5}$, and (iv) the structure $S_5$ with Ir occupancies $d^4$-$d^4$.}
\label{fig:vacrixs}
\end{figure}

\begin{figure}[htpb]
\includegraphics[angle=0,width=\linewidth]{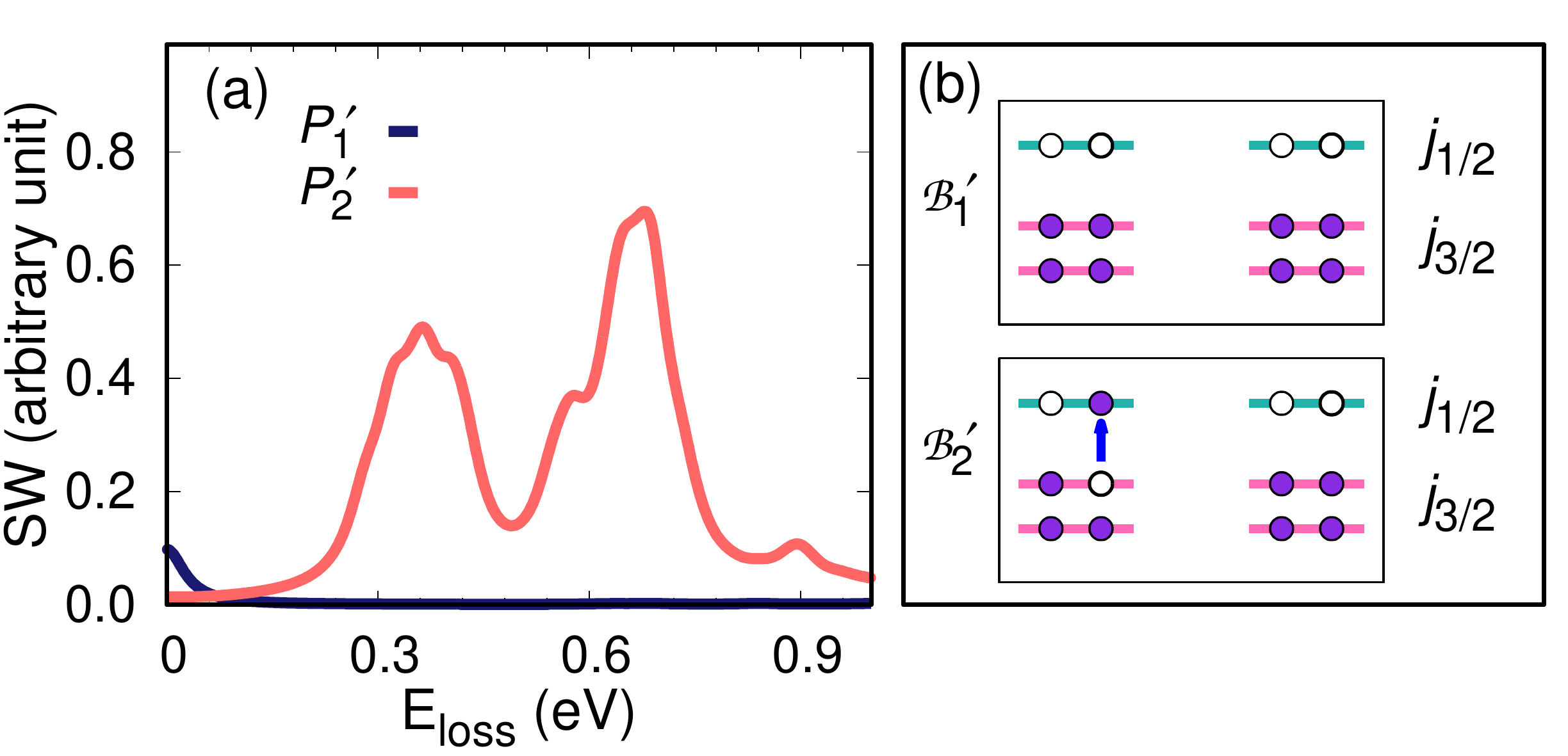}
\caption{(a) Spectral weight [Eq.~\ref{eq:pes}] of various states obtained
from two-site cluster calculations for the $S_4$ structure with Ir occupancies $d^4-d^4$. $P^{\prime}_1$ indicates the ground states, and $P^{\prime}_2$ is local excitons for $d^4$. (b) Schematic diagrams of the lowest-energy subspaces $\mathcal{B}^{\prime}_i$ ($i = 1, 2$) for $d^4-d^4$.}
\label{fig:subspaceS4}
\end{figure}

\begin{table*}[tpb]
\caption {The magnetic interactions in meV for the four structures obtained by exact diagonalization on two-site cluster employing $U$ = 1.7 eV, $J_{\rm H}$ = 0.3 eV, and $\lambda = 0.4$ eV.}
\centering\def\arraystretch{1.1}
\label{tab:hopmag}
\begin{ruledtabular}
\begin{tabular}{lrrrrrrrr}
Structures                 &\multicolumn{2}{c}{$S_1$}           & \multicolumn{2}{c}{$S_2$}     & \multicolumn{2}{c}{$S_3$}  & \multicolumn{2}{c}{$S_4$}           \\
Bonds          & $Z$                & $X$($Y$)            & $Z$      & $X$($Y$)                 & $Z$          & $X$($Y$)            & $Z$     & $X$($Y$)           \\
$J_1$                & 11.4             & 13.7            & 7.3   & 6.9                   &  21.7     &  7.1         & 6.1      & 2.1                 \\
$K_1$                &-6.6              & -5.8            &-9.8   &-10.8                  &  0.7      &  -2.5        & -10.3    & -2.5                 \\
$\Gamma_1$           & 1.6              &  3.5            & 0.8   & 1.5                   &  9.0      &  2.1         & 10.3     & -0.2                 \\
$\Gamma_1^{\prime}$  & -2.1             & -0.7            &-3.6   &-2.8                   & -2.3      &  -0.9        & -1.4     & 2.2                 \\
\end{tabular}
\end{ruledtabular}
\end{table*}

We consider now the effect of Ag vacancies [Figs.~\ref{fig:structure} (c) and~\ref{fig:structure} (d)] in the system. We observe that the consideration of Ag vacancies as introduced in the structures $S_4$ and $S_5$ enhances $t_2$ in the neighboring Ir, as shown in Table~\ref{tab:hop} and induces different occupations of Ir ($d^{4.5}$ and $d^{5}$
in the case of the $Z$-bond).
To take this into account, we performed cluster calculations for the
cases Ir-Ir  $d^5$-$d^5$,  $d^{4.5}$-$d^{4.5}$, and $d^4$-$d^4$ using the hopping parameters of the $Z$-bond in the $S_3$, $S_4$, and $S_5$ structures, respectively. The calculated JDOS and RIXS are displayed in Fig.~\ref{fig:vacrixs}. We observe that RIXS calculated with the
$d^{4.5}$-$d^{4.5}$ cluster shows peaks around 0.06, 0.35, 0.58, 0.8, and 0.9 eV while in the $d^4$-$d^4$ cluster the peaks are around 0.27,  0.39, 0.56, 0.72, and 0.94 eV. Of special importance for both cases, is that they seem to generate an important contribution in the energy region where the B peak in RIXS was reported. Although the ratio of Ir $d^4$ and $d^5$ is expected to be small as described in Ref.~\cite{Torre2021}, the local occupancies $d^{4.5}$-$d^{4.5}$ and $d^4$-$d^4$  in our calculations may still have a measurable contribution to produce a peak comparable to the B peak in experiment.
A smaller ratio of $d^{4.5}$/$d^5$ and $d^4$/$d^5$ in experiment as the one
assumed in our calculations could also explain the smaller magnitude of the
reported B peak in experiment than in our calculation. Furthermore, in the region of 0.4 and 0.8 eV [see Fig.~\ref{fig:vacrixs}],  the comparison of the $d^5-d^5$ case to the
case of introducing Ag vacancies ($d^{4.5}-d^{4.5}$) shows some spectral shifts, which may also explain the slight discrepancies in this region between the experimental observations and the calculations from the stoichiometric case. Further supercell investigations studying
more possible vacancy concentrations and the effect of vacancy-vacancy interactions,
may help in the future to obtain a final picture.

To understand the origin of the B peak, we calculate the dominant spectral weight $P_i{^\prime} (i = 1, 2)$ for $d^4$-$d^4$ clusters in the corresponding energy region. The results in Fig.~\ref{fig:subspaceS4} show that the dominant contribution to peak B is a single exciton via {\it on-site} $j_{3/2}\rightarrow j_{1/2}$ in $d^4$.

\section{Magnetic interactions}\label{sec:mag}

The magnetic interactions for \ALIO~displayed in Table.~\ref{tab:hopmag} were estimated by exact diagonalization in a two-site cluster
of the corresponding multiorbital 
Hubbard model including  spin-orbit coupling interactions~\cite{Winter2016,Winter2017}. The exchange parameters are calculated with
the same parameters as in the previous section, namely $U$ = 1.7 eV, $J_H$ = 0.3 eV, and $\lambda$ = 0.4 eV~\cite{Winter2016, Yamaji2014}. For the experimental structure $S_1$, the Heisenberg exchange interactions are the dominant ones. Averaging the interactions of the $X$ ($Y$) and $Z$ bonds, we obtain ($J_1$, $K_1$, $\Gamma_1$, $\Gamma_1^{\prime}$) $~\sim$ (12.9, -6.1, 2.9, -1.2) meV, leading to Neel AFM magnetic configurations following the classical calculations in Ref.~\cite{Winter2016}.
For the experimental structure $S_3$, $J_1$ is dominant and the anisotropic bond interactions are quite different. The ground state is a Neel AFM magnetic configuration.
For the relaxed structure $S_2$, 
there is less anisotropy between the $Z$ and $X$/$Y$ bonds and the average interactions are ($J_1$, $K_1$, $\Gamma_1$, $\Gamma_1^{\prime}$) $~\sim$ (7.0, -10.5, 1.3, -3.1) meV, leading to the experimentally observed spin-spiral order with the $q$ vector around 0.42 along the a direction, close to 0.32 of {\LIO}. The corresponding Weiss constant is $\Theta^{ab}_0 = -\frac{3}{4k_B}[J + \frac{1}{3} K - \frac{1}{3} (\Gamma + 2\Gamma^{\prime}) = -74.9$ K 
and $\Theta^{c^\ast}_0 = -\frac{3}{4k_B}[J + \frac{1}{3} K  + \frac{2}{3} (\Gamma + 2\Gamma^{\prime})] = -34.2$ K~\cite{Li2021}.
Comparing the values for the $S_2$
structure with {\LIO} ~($J_1$, $K_1$, $\Gamma_1$, $\Gamma_1^{\prime}$) $~\sim$ (-2.7, -8.6, 8.9, -0.6) meV~\cite{Winter2016},
$J_1$ and $\Gamma_1^{\prime}$ are strongly enhanced while $\Gamma_1$ is reduced. 
We also calculated the
second and third neighbor interactions for $S_2$  and the averaged results are ($J_2$, $K_2$, ${\Gamma}_2$, ${\Gamma}_2^{\prime}$, D, $J_3$) $\sim$ (0.5, -0.5, 0.7, 0.4, 0.6, 1.3) meV, which are much smaller than for {\LIO} ~\cite{Winter2016}. For the structure $S_4$ with Ag impurities, the exchange parameters are significantly changed due to the modified hoppings. 

\section{Summary}\label{sec:sum}
In this work we investigate the electronic and magnetic properties of the intercalated honeycomb iridate {\ALIO} by a combination of density functional theory and exact diagonalization of Hubbard models on finite clusters. We show that the magnetism of this system is well captured in terms of  a  localized relativistic $j_{\rm eff}$ = 1/2 basis.
We find that the reported resonant inelastic x-ray scattering spectra can be reproduced 
if Ag vacancies, which introduce both  different Ir filling and modified hybridizations, are assumed. Our results clarify the important role of impurities in intercalated Kitaev candidates. Other systems such as  Ba$_3$CeIr$_2$O$_9$~\cite{Revelli2019} may also need to
invoke the presence of impurities to fully explain  RIXS data. Finally, our magnetic models for the stoichiometric
structure reproduce the experimental observed spin spiral order.

%\newpage
\acknowledgments
We thank Fazel Tafti, Faranak Bahram and Kemp Plump for fruitful discussions. Y.L. acknowledges support by the National Natural Science Foundation of China (Grant No.\ 12004296) and the China Postdoctoral Science Foundation (Grant No.\ 2019M660249). R.V. acknowledges support by the Deutsche Forschungsgemeinschaft (DFG, German
Research Foundation) for
funding through Project No. TRR 288 --- 422213477 (Projects No. A05 and No. B05).

\bibliography{ref}

\appendix
\section{JDOS and RIXS results for the three structures within DFT and ED} \label{app}

\begin{figure}[tpb]
\includegraphics[angle=0,width=\linewidth]{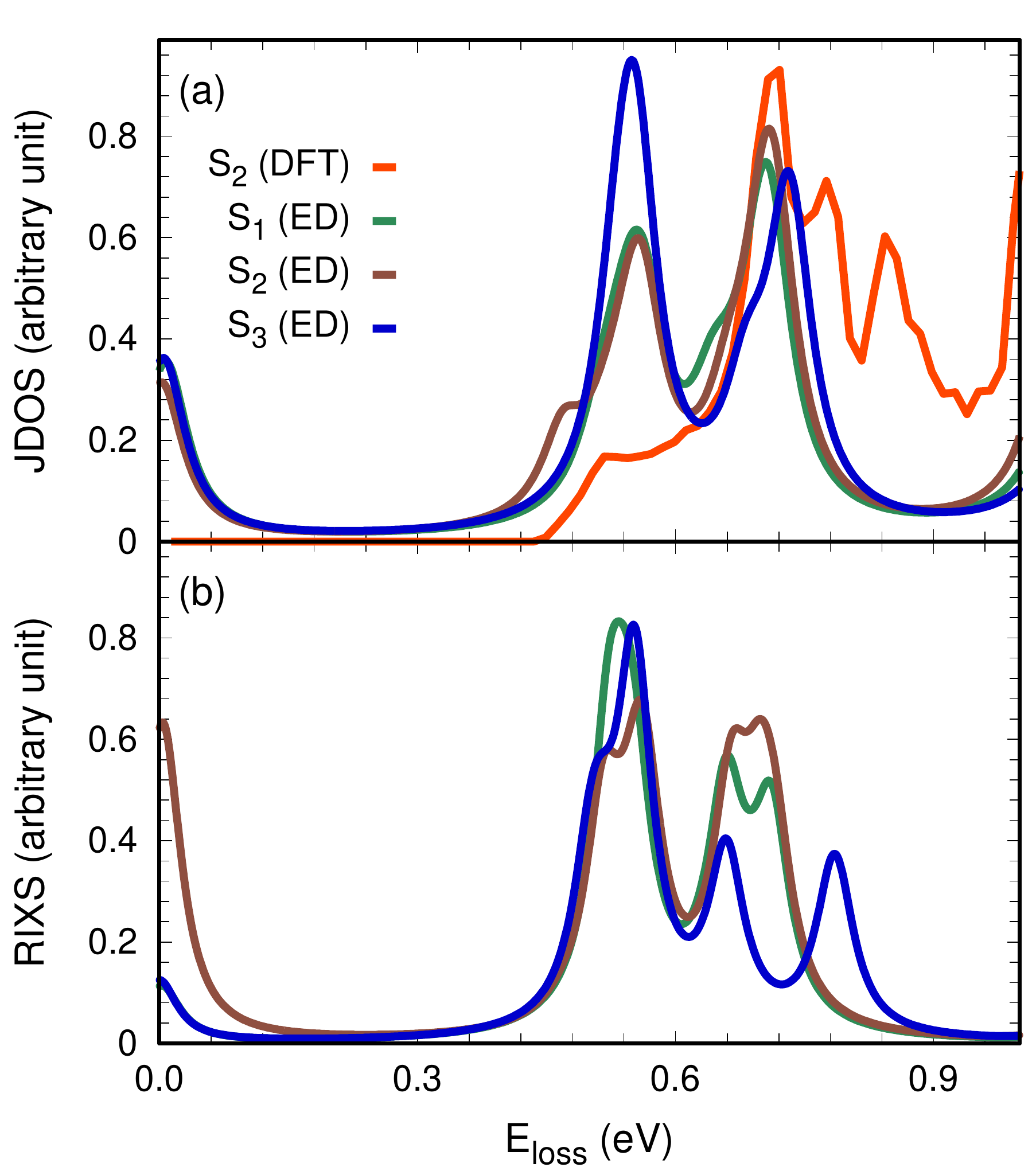}
\caption{(a) JDOS  for $S_2$ obtained from DFT [$S_2$(DFT)] and
from two-site ED [$S_2$(ED)] in comparison to the ED results for $S_1$ and $S_3$.  (b) RIXS results
for $S_1$-$S_3$ structures.
}
\label{fig:jdosrixsapp}
\end{figure}

In Fig.~\ref{fig:jdosrixsapp} (a) we show
 the JDOS calculated (i) with WIEN2k within GGA+SO+$U$ and the stripy magnetic order for the $S_2$ structure and (ii) via exact diagonalization  for the two-site cluster ($Z$-bond) 
for the structures $S_1$, $S_2$, and $S_3$. The corresponding RIXS spectra is
shown in Fig.~\ref{fig:jdosrixsapp} (b). 
The DFT versus ED comparison for JDOS shows that DFT cannot fully capture the effects originating from correlations beyond GGA+SO+$U$, which are expected to be relevant when analyzing electronic excitations. The DOS and RIXS for the $S_1$, $S_2$, and $S_3$ structures indicate that the peaks C and D are robust for the three structures, but E only appears in the structure $S_3$.

\end{document}